\documentclass[prb,aps,graphicx,color,float]{revtex4}
\usepackage{epsfig}
\usepackage{bm}

\begin{document}

\preprint{}

\newcommand{\stef}[2]{$\blacktriangleright${\sc round #1:}{\em #2}$\blacktriangleleft$}

\title{Shot Noise in Anyonic Mach-Zehnder Interferometer}

\author{ D. E. Feldman$^1$, Yuval Gefen$^2$, Alexei Kitaev$^{3,4}$, K. T. Law$^1$, and Ady Stern$^2$}
\affiliation{$^1$Department of Physics, Brown University, Providence,
Rhode Island 02912, USA\\
$^2$Condensed Matter Physics Department, Weizmann Institute of Science, Rehovot,
76100 Israel\\
$^3$California Institute of Technology, Pasadena, California 91125, USA\\
$^4$Microsoft Project Q, University of California, Santa Barbara,
California 93106, USA}

\begin{abstract}

We show how shot noise in  an electronic Mach-Zehnder
interferometer in the fractional quantum Hall regime probes the
charge and statistics of quantum Hall quasiparticles. The
dependence of the noise on the magnetic flux through the
interferometer allows for a simple way to distinguish Abelian from
non-Abelian quasiparticle statistics. In the Abelian case, the
Fano factor (in units of the electron charge) is always lower than
unity. In the non-Abelian case, the maximal Fano factor as a
function of the magnetic flux exceeds one.

\end{abstract}

\pacs{73.43.Jn, 73.43.Cd, 73.43.Fj}

\maketitle

\section{Introduction}

Among the most fascinating features of the quantum Hall effect
(QHE) are the fractional charge and statistics of elementary
excitations. The charge of a quasiparticle is a simple fraction of
an electron charge. In most QHE states the quasiparticles are
Abelian anyons, i.e., they accumulate nontrivial phases when
encircle each other. For some filling factors, even more
interesting non-Abelian statistics was
predicted~\cite{MooreRead,nw}. When a single non-Abelian anyon
makes a full circle around another (identical) one, this not only
modifies the phase of the  wave function, but can also result in a
change of the quantum state of the quasiparticle system. Such a
property makes non-Abelian anyons promising for fault-tolerant
quantum computation \cite{Kitaev97,kitaev}.

Detecting fractional charge and statistics proved to be a
difficult task. Fractional  charges were eventually observed in
shot noise experiments at Weizmann Institute and
Saclay~\cite{weizmann-G}. No non-ambiguous  observation of the
statistics of {identical} fractional quasiparticles has been
reported so far  in spite of several theoretical proposals for
detecting Abelian \cite{cfksw,HBT,kane,lfg,gss} and non-Abelian
\cite{FNTW,DFN,BKS,SH,gss,chamon,fk} anyons, and important  efforts in this
direction \cite{Goldman,Rosenow}. The set-up \cite{weizmann-G},
which was  successful for measuring fractional charges via shot
noise, cannot be applied to the problem of fractional statistics
since no quasiparticle braiding occurs in such a device. 

Recently a different device, an electronic Mach-Zehnder interferometer,
has been designed and  fabricated at Weizmann Institute \cite{MZ},
Fig. 1. Quasiparticle braiding is possible in that configuration.
As we show below, shot noise in the Mach-Zehnder interferometer
contains information about both charge and statistics of the
elementary excitations. In particular, it allows for a simple and
physically transparent way to distinguish Abelian from non-Abelian
quasiparticles.

Anyon statistics is defined in terms of quasiparticle braiding. 
Hence, interference experiments may be a useful tool for experimentally observing it. 
It has been proposed to employ electronic Fabry-Perot and Mach-Zehnder interferometers for
probing Abelian \cite{cfksw,lfg} and non-Abelian \cite{FNTW,DFN,BKS,SH,fk} statistics. The
Fabry-Perot approach is simple and transparent: it is based on quasi-particles moving along the edge encircling quasi-particles localized in the bulk. This approach is however sensitive to fluctuations of the number of the trapped
quasiparticles \cite{kane,BKS,newreview}
which can tunnel to and from localized states inside the interferometer. In the absence of these fluctuations the interference pattern that results from the lowest-order back-scattering amplitude shows clear and unambiguous signatures of Abelian and non-Abelian statistics. When such fluctuations are present and are slow compared to the scale of the particle time-of-flight through the interferometer, lowest order interference is wiped out. The remaining signal, originating from multiple reflections, shows less striking signatures in the visibility of the interference. An additional signature is seen in this case in current noise\cite{gss}.
The Mach-Zehnder interferometer is not sensitive to these slow fluctuations \cite{lfg,fk}, and shows signatures of Abelian and non-Abelian statistics in the visibility of the interference and in the $I-V$ characteristics.  Fast fluctuations suppress the interference in the two geometries.
In both cases, the experimental outcome depends on unknown nonuniversal microscopic
parameters.
Obviously, it would be desirable to find a simpler procedure based on a single 
measurement of a universal quantity.
We show below that this can be accomplished in 
a shot noise experiment with an anyonic Mach-Zehnder interferometer. The
information about statistics is contained
in a single universal quantity: the maximal Fano factor at low temperatures.

Nonequilibrium noise is an important experimental tool in mesoscopic physics.
We define it as the Fourier transform of the current-current correlation function \cite{footnote},

\begin{equation}
\label{noise}
S(\omega)=\frac{1}{2}\int_{-\infty}^{\infty}\langle \hat I(0) \hat I(t) + \hat I(t) \hat I(0)\rangle \exp(i\omega t) dt.
\end{equation}
It is often convenient to express the noise in terms of an
effective   charge  $e^*(\omega)$, according to $S=e^*\langle \hat
I\rangle$. The dimensionless ratio $e^*(\omega)/e$ is known as the
Fano factor and  depends on the temperature, voltage and
frequency, disorder and interaction strength. In some situations
$e^*$  is universal and equals the carrier charge. This property
was used for probing fractional  charges in Ref.
\onlinecite{weizmann-G}. Shot noise in the Mach-Zehnder
interferometer in the integer QHE has recently attracted much
attention \cite{mznoise}. Below we show  that the physics of noise
is even richer  in the fractional QHE regime. We find that the
Fano factor in the Mach-Zehnder interferometer depends on
frequency. At high frequencies the zero-temperature Fano factor is
equal to the dimensionless quasiparticle charge. At low
frequency, the Fano factor depends on the magnetic flux. The
maximum value of the zero-temperature Fano factor is always below
$1$  in Abelian systems and can exceed $1$ for non-Abelian anyons.

Fig. 1 shows the structure of the interferometer~\cite{MZ}. In the absence
of inter-edge tunneling, charge propagates along two chiral edges
(S1 to  D1 and  S2 to D2); the source S1 is kept at a finite
voltage $V$ with respect to S2. Quasiparticles can tunnel between
the edges at two point contacts QPC1 and QPC2.
The current that flows between the edges (i.e., from S1 to D2)
depends on  $V$,   the magnetic flux $\Phi$ through the loop
A-QPC2-B-QPC1-A, and the tunneling amplitudes, $\Gamma_1$ and
$\Gamma_2$, at the two quantum tunneling contacts. We assume that
the tunneling amplitudes are small. Hence,  the mean time  between
two consecutive quasiparticle tunneling events is much larger
than the duration of  such an  event. We also assume that the
leads fully absorb edge excitations. For a system of bosons or
fermions such an assumption implies that  tunneling events are
independent. In contrast, we will see that the probability of an
anyon to tunnel  is affected by the history of  previous tunneling
events.

Below we calculate the noise for the tunneling current between the edges.
We will distinguish the high- and low-frequency regimes. In the low-frequency regime,
the frequency $\omega$ is less than $1/\tau_T=I/q$, where $\tau_T$ is the typical time interval 
between consecutive
tunneling events, $I$ the average current and $q$ the quasiparticle charge. In the high-frequency regime,
$1/\tau_J>\omega>1/\tau_T$, where $\tau_J$ is the time of one tunneling event. The time 
$\tau_J\sim\tau_{\rm travel}+\tau_{\rm uncertainty}$ can be estimated as the sum of two contributions. 
The quasiparticle travel time between the point contacts, $\tau_{\rm travel}$,
is of the order of
$\sim L/v$, where $L$ is the interferometer size and $v$ the quasiparticles velocity along the edges. The time 
$\tau_{\rm uncertainty}\sim \hbar/{\rm max}(qV,k_BT)$, where $V$ is the voltage bias and $T$ the temperature,
comes from the uncertainty of the energy of tunneling quasiparticles.

Most of this paper focuses on the low-frequency noise at zero temperature and on the finite temperature noise at arbitrary frequency.
The zero-temperature high-frequency limit is much simpler. 
At high frequencies, the contribution to Eq. (\ref{noise}) from large times
$t\gg 1/\omega$ is suppressed by the oscillating exponential
factor. The contribution from short times $t\sim \tau_J$ can be found
by substituting $\omega=0$. One gets $S=\langle\hat I(0)
\int^{\tau_J}_{-\tau_J} dt\hat I(t)\rangle=e^*I$, where the Fano factor
$e^*=\int_{-\tau_J}^{\tau_J}I(t)dt$ equals the charge transmitted during
one tunneling event. Thus, in the high frequency limit, $e^*$ 
is the quasiparticle charge $q$. This result is similar to the
low-frequency behavior of the noise in the geometry with one
tunneling contact \cite{kf94} and holds for any quasiparticle statistics.
In contrast, the low-frequency noise in the Mach-Zehnder
interferometer shows a unique behavior which will allow detecting fractional statistics.

The article has the following structure. Section II contains the main results of the paper.
It focuses on the low-frequency noise
at $T=0$ in two types of QHE states: Laughlin states
with filling factors $\nu=1/m$ ($m$ odd) and the Pfaffian
state~\cite{MooreRead,nw} at filling factor $\nu=5/2$. 
The technical approach of Section II is rather simple and its results can be understood from
a simple qualitative picture. The rest of the paper
focuses on the finite temperature noise for arbitrary fractional statistics and is more technical. 
The method is based on the algebraic theory of anyons, Ref. \onlinecite{kitaev}. 
Section III contains the calculation of the tunneling probabilities. In Section IV we find the tunneling current.
In Section V we calculate the zero-frequency noise at arbitrary temperatures. Section VI discusses the high-frequency noise with the emphasis on the limit of low voltages and finite temperatures, i.e. the linear response regime. In that regime the noise and conductance are related by the 
Nyquist formula. We find that in contrast to the shot noise, the thermal Nyquist noise is independent of the magnetic flux 
through the interferometer for any fractional statistics with nondegenerate braiding.

\section{Shot noise at zero temperature}

In this section we consider the Laughlin and Pfaffian states.
We first briefly review the anyon statistics in these two states and then calculate
the noise.

\subsection{Fractional statistics in Laughlin and Pfaffian states}

The quasi-hole charge in a Laughlin state with the filling factor $\nu$ is $q=\nu e$. When
one quasi-hole encircles  $n$ others counterclockwise, the wave
function accumulates the statistical phase
\begin{equation}
\label{Abelian_phase_add}
\label{abelian} \theta_s=2\pi \nu n.
\end{equation}
Since the phase $\theta_s$ is defined ${\rm mod}~2\pi$, there are
$r=0,\dots,1/\nu-1$  ``equivalence classes"  which correspond to
$1/\nu$ physically distinct statistical phases $\theta_s$. A
configuration of $n$ quasi-holes belongs to the 
$r$-th class, $r=n\  {\rm mod}~1/\nu$.

In the $\nu=5/2$ Pfaffian liquid, quasi-holes carry charge $e/4$.
Due to the non-Abelian statistics, the quantum state of a system
of several charge-$e/4$ quasiholes is not uniquely determined by
their coordinates. One needs to specify additional quantum
numbers, most importantly, the topological charge (for a review,
see Ref.~\onlinecite{kitaev}.) In the $\nu=5/2$ state the latter
takes on one of  three values~\cite{SH}: $1$ (``vacuum''),
$\epsilon$ (``fermion''), and $\sigma$ (``vortex''). In a system
with an odd number of quasi-holes $n$, the topological charge is
always $\sigma$. At even $n$ two topological charges, $1$ and
$\epsilon$, are possible. When two sets of quasi-holes merge, the
topological charge of the composite system is given by the fusion
rules: $\epsilon\times\epsilon=1$,\,
$\epsilon\times\sigma=\sigma$,\, $\sigma\times\sigma=1+\epsilon$.
In a process in which a quasi-hole encircles a set  of $n$
quasi-holes whose topological charge is $\alpha$, the wave
function picks up a statistical phase which depends on $n$,
$\alpha$ and the topological charge $\beta$ of the whole system
consisting of  $(n+1)$ quasi-holes. We denote this statistical
phase by $\phi_{ab}$, where $a=(n,\alpha)$ and $b=(n+1,\beta)$
label superselection sectors \cite{kitaev}. It equals
\begin{equation}
\label{2}
\phi_{ab}=n\pi/4+\phi'_{\alpha\beta},
\end{equation}
where the non-Abelian part $\phi'_{\alpha\beta}$ satisfies the
following rules: $\phi'_{1 \sigma }=0$,
$\phi'_{\epsilon\sigma}=\pi$, $\phi'_{\sigma 1}=-\pi/4$ and
$\phi'_{\sigma\epsilon}=3\pi/4$. As we will see this information
determines the quasi-hole tunneling rates. The statistical phase
factor $\exp(i\phi_{ab})$, Eq.~(\ref{2}), is unchanged if the
state  $a=(n,\alpha)$ is fused  with an electron whose
superselection sector is $(4,\epsilon)$. Thus, the superselection
sectors form 6 equivalence classes, which are characterized by
$(n\bmod 4,x)$, where $x=\sigma$ for odd $n$ and either  $1$ or
$\epsilon$ for even $n$.

Due to the energy gap for bulk excitations, the low-energy physics of the 
Mach-Zehnder interferometer is determined by the
edges. Hence, the Hamiltonian is
\begin{equation}
\label{3}
\hat H=\hat H_{\rm edge}
+[(\Gamma_{1}\hat X_1 + \Gamma_{2}\hat X_2) + H.c.],
\end{equation}
where $\hat H_{\rm edge}$ is the Hamiltonian of the two edges and
the operators $\hat X_1$, $\hat X_2$ correspond to the quasi-hole
transfer from outer edge 1 to inner edge 2 at QPC1 and QPC2,
respectively. Strictly speaking, the Hamiltonian must include operators
which describe tunneling of the objects with all possible electric 
and topological charges. We take into account only the quasi-hole tunneling operators since
they are most relevant \cite{foot1}.
In the limit of small tunneling amplitudes
$\Gamma_1$ and $\Gamma_2$, the probability of a tunneling event
from edge 1 to edge 2 includes four contributions: two
proportional to $|\Gamma_{1,2}^2|$ (independent of the flux $\Phi$
through the interferometer) and two  $\propto \Gamma_1\Gamma_2^*$
and $\propto \Gamma_1^*\Gamma_2$ (flux dependent). The last two
describe the interference of quasi-hole trajectories S1-QPC1-B-D2
and S1-A-QPC2-D2 and include   the Aharonov-Bohm phase $2\pi
q\Phi/[e\Phi_0]$. Here   $q$ is the quasiparticle charge and
$\Phi_0=eh/c$ the flux quantum.

Furthermore and central to our discussion is the  dependence of
these interference terms  \cite{lfg,fk} on the topological charge
of edge 2. In the case of the Laughlin states the tunneling
probability depends on the statistical phase Eq. (\ref{abelian}),
where $nq$ is the total charge transferred between the edges
(negative $n$ corresponds to tunneling events from edge 2 to edge
1). In the Pfaffian state these terms contain the phase 
(\ref{2}), where $n$ has the same meaning as in the Abelian case,
and $\alpha$ and $\beta$ are the initial and final topological
charges of edge 2. In both cases a given tunneling event modifies
the tunneling rate of the subsequent event. Quasiparticle
tunneling between the edges can be viewed as a Markov process
\cite{lfg,gss,fk}.

\subsection{ Abelian  Laughlin case, $\nu=1/m$} 

The tunneling rates were
computed in Ref. \onlinecite{lfg} (see also Section III). Transitions are possible for
the states connected by arrows in Fig. 2a.  At zero temperature,
quasi-holes tunnel only from the edge with higher chemical
potential (edge 1) to the edge with the lower potential, i.e.,
only in the direction of the arrows. The precise expressions for
the transition rates depend on the edge state model \cite{lfg}. We
are concerned only with manifestations of fractional statistics
and discuss below only those properties of the transition rates
which depend on statistics. The transition rate depends \cite{lfg}
on the number of the previous tunneling events ${\rm mod}~m$.
Thus, at zero temperature and low voltage, there are $m$ different
tunneling rates \cite{lfg}
\begin{equation}
\label{prob_Abelian}
p_k=r_0(V)\{|\Gamma_1|^2+|\Gamma_2|^2+[u(V)\Gamma_1^*\Gamma_2\exp(2\pi
i\nu[\Phi/\Phi_0+k])+c.c.]\},
\end{equation}
where $k=0,\dots,m-1$ and the coefficients $r_0$ and $u$ are
expressed  via the matrix elements of the operators $\hat
X_{1,2}$. The above perturbative expressions for the rates are
valid \cite{lfg,fk} for such small $\Gamma_{1,2}$ that $p_k\ll
qV/h$ for all $k$. The coefficient $u$ depends on the voltage and
the distance between the tunneling contacts. For $qVL/hv\ll 1$ it
follows that $u\rightarrow 1$ (here $L$ is the interferometer size
and $v$ is the edge velocity). If also
$|\Gamma_1|\approx|\Gamma_2|$,  it is possible to tune the
magnetic flux in such a way that one of the rates becomes much
smaller than all others.


Let us now calculate the Fano factor.
The average square of the time
between two subsequent tunneling events is
$\overline{t_k^2}=2/p_k^2$. The noise (\ref{noise}) equals
$S=\overline{\delta Q^2(t)}/t,$ where $\delta Q(t)$ is the
fluctuation of the charge $Q(t)$ transmitted during the time $t\gg
1/p_k$. The Fano factor reads as
\begin{equation}
\label{Fano} e^*=S/I=\overline{\delta Q^2(t)}/\overline{Q(t)}.
\end{equation}
The charge transmitted during $N/\nu\gg 1$ tunneling events is
$Ne$. The average time $t_N$ needed for $N/\nu$ tunneling events
is $\bar t_N=N\sum_{k=0}^{1/\nu-1}\frac{1}{p_k}$. The time
$t_N=\bar t_N+\delta t$ fluctuates. One can estimate $\overline
{Q(\bar t_N)}$ as $Ne$ with the accuracy of $O(\delta t/t_N)$.
Such accuracy is insufficient for the calculation of
$\overline{\delta Q^2(\bar t_N)}$. For the sake of evaluating the
second moment of the current one can either calculate the
fluctuations of $Q$ during
 $\bar t_N$ or assume a uniform current with fluctuations of the time interval
$\delta t$. We thus need to find the fluctuation $\overline{\delta
t^2}$.
From the expression for $\overline{t_k^2}$ we find
  $\overline{\delta
t^2}=N\sum_k \frac{1}{p_k^2}$. Now we can estimate ${Q(\bar
t)}\approx Ne-\delta t I={\rm const} -\delta t Ne/\bar t_N$.
Hence, $\overline{\delta Q^2}=I^2\overline{\delta t^2}$. Finally,
\begin{equation}
\label{Laughlin} e^*=e\frac{\sum_{k=0}^{1/\nu-1}
1/p_k^2}{(\sum_{k=0}^{k=1/\nu-1} 1/p_k)^2}
\end{equation}
Since the rates $p_k$ depend on the magnetic flux $\Phi$, the
low-frequency Fano factor depends on flux as well, in contrast to
the one-point-contact geometry \cite{kf94}. From the inequality of
arithmetic and quadratic means we find that $e^*\ge \nu e$. The
maximum value of $e^*$ is the electron charge $e$, and it is
obtained when one of the transition rates is much smaller than all
others (see discussion below).

\subsection{Shot noise in non-Abelian Pfaffian state}  

In the Pfaffian state there are
8 transition rates \cite{fk} (see Fig. 2b; a general approach to the calculation of transition rates
is discussed in Section III). At zero temperature
and low voltage four of them have the form \cite{fk}
\begin{equation}
\label{prob} p_k=r_0\{|\Gamma_1|^2+|\Gamma_2|^2+[u\Gamma_1^*
\Gamma_2\exp(\pi i[\Phi/\Phi_0+k]/2)+c.c]\},
\end{equation}
where $k=0,1,2,3$. The four remaining rates are given by  $p_k/2$
(Fig. 2b). As in the Laughlin state, if
$|\Gamma_1|\approx|\Gamma_2|$ and $u\approx 1$ it is possible to tune the magnetic
flux in such a way that one of the rates becomes much smaller than
all others. The condition $u\approx 1$ corresponds to close point contacts, i.e.
the distances between the point contacts along both edges must be shorter than all other
scales: the edge lengths; the scale $hv/qV$, controlled by the voltage bias; and $hv/k_BT$, where $T$ is the temperature.

Using the same approach as in the Abelian case, one can rewrite
Eq. (\ref{Fano}) as
\begin{equation}
\label{Fano1} e^*=e{\overline{\delta t^2}}/{(\bar t)^2},
\end{equation}
where $\bar t$ is the average time needed for 4 consecutive
tunneling events and $\delta t$ is the fluctuation of that time.
Since one electron charge is transfered between the edges during 4
tunneling events, the current reads as
\begin{equation}
\label{current} I=e/\bar t.
\end{equation}
A straight-forward calculation yields
\begin{equation}
\label{time} \bar t=2[{\sum_{k=0}^3p_k/p_{k+1}+4}]/[{\sum_{k=0}^3
p_k}];
\end{equation}
and
\begin{equation}
\label{fl} \overline{\delta
t^2}=\left[\frac{4}{\sigma}\sum_{k=0}^3\frac{p_k+p_{k+1}}{p_{k+1}^2}-
\frac{4}{\sigma^2}\left(\sum_{k=0}^{3}\frac{p_k}{p_{k+1}}\right)^2\right]
\end{equation}
where $\sigma=\sum p_k$ and the convention $p_{3+1}=p_0$ is used.
The calculation is based on Fig. 2b. In the space of states, there are four possible trajectories 
which begin and end in point $(-1,\sigma)$. $\bar t$ and $\overline{t^2}$ are combinations of
the contributions from the 4 trajectories weighted according to their probabilities. For example,
the probability of the trajectory 
$(-1,\sigma)\rightarrow (0,1) \rightarrow (1,\sigma) \rightarrow (2,1)\rightarrow (-1,\sigma)$
is $p_3/(p_1+p_3)\times p_0/(p_0+p_2)$.
In order to make Eqs. (\ref{time},\ref{fl}) more compact and
symmetric, we used the fact that $p_0+p_2=p_1+p_3$ which is
evident from Eq. (\ref{prob}).

Equations (\ref{Fano1}-\ref{fl}) have an interesting feature: in
contrast to the Abelian case, Eq. (\ref{Laughlin}), the Fano
factor can exceed one electron charge. A convienient reference point corresponds to such magnetic 
flux $\Phi$ that the current (\ref{current}) is minimal.  At $u\approx 1$ and
$\Gamma_1\approx\Gamma_2$, the current (\ref{current},\ref{time}) is
minimal for $\Phi=(4n-k+2)\Phi_0$, in which case the tunneling
rate $p_k$, Eq. (\ref{prob}), is close to zero. Assuming $k=0$ one gets $p_0\ll
p_1,p_2,p_3$. Then Eqs. (\ref{prob}) and (\ref{Fano1}-\ref{fl}) imply
that $e^*=3e>e$. The maximal Fano factor is achieved at a different value of the magnetic flux.
It can be found numerically from
Eqs. (\ref{Fano1},\ref{time},\ref{fl}). In the limit $u\approx 1$ and $\Gamma_1\approx\Gamma_2$, 
the maximal effective charge equals $e^*_{\rm max}\approx 3.2e$.
In contrast, the maximal value of the Fano factor
of the Laughlin state, Eq. (\ref{Laughlin}), is $e^*=e$.

The difference between the two cases provides a way to distinguish
between Abelian and non-Abelian statistics from shot noise. In
fact, this difference may be understood from Fig. 2 without
detailed calculations. Both in the Abelian and non-Abelian cases
the minimal current corresponds to $p_0\approx 0$. Transport in a
Laughlin state can be described in the following way: typically
the system stays in state 0 (Fig. 2a) for a long time after which
it tunnels to state 1. From state $1$ it quickly goes through the
series of states $2, 3, ..., m-1$ and back to $0$. Then the system
stays in state 0 again for a long time. Thus, charge is typically
transmitted in ``bursts" of $m$ quasiparticles, carrying altogether
one electron charge. This gives $e^*=e$. In the non-Abelian case,
Fig. 2b, the transport also occurs rapidly between prolonged stays
in state $(0,1)$. However, the average charge transmitted between
two such stays exceeds an electron charge due to the existence of
a ``bypass road''
$(-1,\sigma)\rightarrow(0,\epsilon)\rightarrow(1,\sigma)\rightarrow(2,\epsilon)\rightarrow(-1,\sigma)$.

\section{Transition rates}

In this section we calculate transition rates for arbitrary fractional statistics at arbitrary temperature $T$ and voltage $V$. We assume that the topological charge of elementary excitations has a finite number of possible values.
There is an infinite number of possible electric charges. We will denote the minimal charge of an anyon as $q>0$.
The quantum dimensions $d_a$, fusion multiplicities $N^b_{ax}$ and statistical phases $\phi_{ab}$ [Eq. (\ref{statphase})] depend only on the topological charges. All these conditions are satisfied for the Laughlin and Pfaffian statistics for an appropriate definition of the topological charge: Each vertex of the graphs in Fig. 2 should be viewed a class of states with the same topological charge, and different topological charges should be ascribed to different vertices.

We will work in the limit of small $\Gamma_1$ and $\Gamma_2$, Eq. (\ref{3}), so that the transition rates
can be found from the lowest, second, order in $\Gamma$'s. Physically this means that at every moment of time there is no more than one quasiparticle between the point contacts. Nevertheless, the transition rates depend on all previous tunneling events through the total topological charge of the edges.

Consider a tunneling event in which a quasiparticle with electric charge $q$ and topological charge $x$ is transfered from the outer edge 1 to the inner edge 2 (Fig. 1). After such an event the topological charge of edge 2 changes. We will denote the initial topological charge of edge 2 as $a$ and the final topological charge as $b$. Possible values of $b$ are determined
by the fusion rules for charges $x$ and $a$. Since the total topological charge of the interferometer must remain at the vacuum value 1, the topological charge of edge 2 is always opposite to the topological charge of edge 1. In what follows we
will use the notation $\bar a$ for the topological charge opposite to $a$.

The tunneling operators $\hat X_1$ and $\hat X_2$, Eq. (\ref{3}), depend on the gauge. The choice of the vector potential $\vec A$
is restricted by the condition that its circulation over the loop A-QPC2-B-QPC1-A (Fig. 1) equals the magnetic flux $\Phi$ through the interferometer. At the most convenient choice of the gauge, the vector potential is nonzero only inside the point contact QPC2. In this case the edge Hamiltonian and the tunneling operator $\hat X_1$ are independent of the magnetic flux and the operator $\hat X_2$ contains the factor $\exp(i\phi_{\rm AB})$, where the Aharonov-Bohm phase
$\phi_{\rm AB}=2\pi i q\Phi/e\Phi_0$ and $\Phi_0=hc/e$ is the flux quantum. 

In addition to the Aharonov-Bohm phase, a particle accumulates a statistical phase $\phi_{ab}$ when it goes
along the loop A-QPC2-B-QPC1-A. Hence, $\hat X_2$ contains an additional factor $\exp(i\phi_{ab})$. For the calculation of
the statistical phase we consider a process in which a particle with topological charge $x$ encircles the hole in the interferometer (Fig. 1) which contains topological charge $a$. The result of this calculation depends on the fusion channel $b$ for the charges $a$ and $x$. 
The phase can be found from algebraic theory of anyons, Ref. \onlinecite{kitaev}, using the
diagrams from Fig. 3. The left diagram equals the topological spin $\theta_b$ while the right diagram equals 
$\theta_a\theta_x\exp(i\phi_{ab})$. Hence,

\begin{equation}
\label{statphase}
\exp(i\phi_{ab})=\frac{\theta_b}{\theta_a\theta_x}.
\end{equation}
Thus, the tunneling operators can be represented in the form 
$\hat X_1=\tilde X_1$, $\hat X_2=\exp(i\phi_{\rm AB}+i\phi_{ab})\tilde X_2$,
where the operators $\tilde X_i$ are independent of the magnetic flux and the topological charges of the edges.

Since a tunneling event involves a change of the topological charge of edge 2, we have to find the probabilities of all possible fusion outcomes $b$. The tunneling operators $\tilde X_i$ create a quasiparticle-quasihole pair,
$x$ and $\bar x$, near one edge inside a point contact and move the quasiparticle $x$ to the opposite edge. Since the topological charge is an integral of motion, the quasiparticle and quasihole fuse to vacuum. We want to find the amplitude of the process in which $x$ and $a$ fuse to $b$. This amplitude is given by the diagram in Fig. 4. The vertices are two operators $F_{ax;i}^b$ and $F_{\bar a\bar x;j}^{\bar b}$  from the fusion space $V_{ax}^b$, Ref. \onlinecite{kitaev}; the indeces
$i$ and $j$ assume values between 1 and the fusion multiplicity $N^b_{ax}$. The diagram, Fig. 4, is well known in  algebraic theory of anyons. In order to apply the expression from Ref. \onlinecite{kitaev} for this diagram, one just
needs to remember that in algebraic theory of anyons, quantum states have norms different from 1. Thus, up to an irrelevant phase, the fusion amplitude equals $f_{ij}=[1/d_a d_x]\sqrt{d_a d_b d_x}\delta_{ij}$, where the square bracket contains
the normalization factor and $d_\alpha$ denote quantum dimensions. Finally, the fusion probability

\begin{equation}
\label{fusionprob}
P^+_{a\rightarrow b}=\sum_{ij}|f_{ij}|^2=N^b_{ax}\frac{d_b}{d_ad_x}.
\end{equation}

The total transition rate equals

\begin{equation}
\label{w+}
w^+_{a+x\rightarrow b}=P^+_{a+x\rightarrow b}u^+_{a+x\rightarrow b},
\end{equation}
where 
$u^+_{a+x\rightarrow b}$
is the transition rate in the fusion channel $b$. We consider fully absorbing leads. Hence, the rate 
$u^+_{a+x\rightarrow b}$ depends on the previous tunneling events only through the factor $\exp(i\phi_{ab})$ in the operator $\hat X_2$. The rate can be found from Fermi's golden rule:

\begin{equation}
\label{u-rates}
u^{+}_{a+x\rightarrow b}=r_{11}^{+}\left(|\Gamma_1|^2+|\Gamma_2|^2\right)
+\left(r_{12}^{+}e^{i\phi_{\rm AB}}e^{i\phi_{ab}}\Gamma_1^*\Gamma_2
+c.c.\right),
\end{equation} 
where $r^+_{\alpha\beta}=\int_{-\infty}^{\infty}\langle \tilde X_{\alpha}^\dagger (t) \tilde X_{\beta}(0)\rangle/\hbar^2$
and the angular brackets denote the average with respect to the thermodynamic state of noninteracting edges at the temperature $T$ and the chemical potential difference $qV$ between the edges. The correlation functions $r^+_{\alpha\beta}$
depend on microscopic details.

At finite temperatures, quasiparticles can also tunnel from internal edge 2 to edge 1 whose electrochemical potential is higher. The respective transition rates 

\begin{equation}
\label{w-}
w^-_{b\rightarrow a+x}=P^-_{b\rightarrow a+x}u^-_{b\rightarrow a+x}
\end{equation}
 can be found in 
exactly the same way as $w^+$. In full analogy with Eq. (\ref{fusionprob}), the diagram from Fig. 4 with interchanged
$b\leftrightarrow\bar a$ and $\bar b\leftrightarrow a$ yields

\begin{equation}
\label{fusionprob2}
P^-_{b\rightarrow a+x}=N^b_{ax}\frac{d_a}{d_bd_x}.
\end{equation}
The transition rate $u^-_{b\rightarrow a+x}$ within the fusion channel 
$b\rightarrow a+x$ is related to $u^{+}_{a+x\rightarrow b}$
by the detailed balance condition

\begin{equation}
\label{dbc}
u^-_{b\rightarrow a+x}=\exp(-qV/T)u^{+}_{a+x\rightarrow b}.
\end{equation}

Equations (\ref{fusionprob}-\ref{dbc}) allow one to easily reproduce Eqs. (\ref{prob_Abelian},\ref{prob}) from the previous section. One just needs to know the quantum dimensions, topological spins and fusion multiplicities. 
In Laughlin states all quantum dimensions and fusion multiplicities are 1. The topological spin of the excitation with the electric charge $ne\nu$ equals $\exp(i\nu n^2)$, where $\nu$ is the filling factor. The topological spins and quantum dimension in the Pfaffian states are listed in Table I (cf. Refs. \onlinecite{kitaev,SH,fk}). 
All fusion multiplicities $N^b_{ax}=1$.

\begin{table}
\vspace{3mm}
\begin{tabular}{|c|c|c|c|}
	\hline
Topological charge &  Electric charge    &   Quantum dimension  & Topological spin     \\
	\hline
          1        & $ne/2$              & 1        & $\exp(i\pi n^2/2)$    \\
$\epsilon$ & $ne/2$    & 1          & $-\exp(i\pi n^2/2)$\\
$\sigma$ & $e/4+ne/2$ & $\sqrt{2}$ &  $\exp(i\pi[2n^2+2n+1]/4)$ \\

	\hline
\end{tabular}
\caption{Quantum dimensions and topological spins for excitations with different electric and topological charges in the Pfaffian state.}
\label{table_I}
\vspace{3mm}
\end{table}

\section{Finite-temperature current}

Now we are in the position to calculate the tunneling current through the interferometer. 
We consider the most general situation when the tunneling of several types of quasiparticles 
with electric charges $kq$ is allowed ($k$ are integers). Our final answer, Eq. (\ref{AA}), reproduces the results of Section II
in the limit of zero temperature for the Laughlin and Pfaffian states.

Let $Q(t)$ be the total
charge of edge 2. Then the current

\begin{equation}
\label{Q/t}
I=\lim_{t\rightarrow\infty}[Q(t)-Q(0)]/t.
\end{equation}
The charge $Q(t)$ can be found from the generating function $q(z,t)=q\sum p_{a,nq}z^n$, where $p_{a,nq}$ is
the probability to find topological charge $a$ and electric charge $nq$ on edge 2. One gets

\begin{equation}
\label{genQ}
Q=\frac{d}{dz}q(z,t)|_{z=1}.
\end{equation}
The probabilities $p_{a,nq}$ obey the kinetic equations

\begin{equation}
\label{kinetic}
\dot p_{a,nq} = \sum_{b,mq} M^{a,nq}_{b,mq}p_{b,mq},
\end{equation}
where the kinetic matrix 

\begin{equation}
\label{Mab}
M^{a,nq}_{b,mq}=\sum_x [w^+_{(b+x),m\rightarrow a,n}+w^-_{b,m\rightarrow (a+x),n}]
-\delta_{ab}\delta_{nm}\sum_{c,l,x}[w^+_{(a+x),n\rightarrow c,k}+w^-_{a,n\rightarrow (c+x),k}].
\end{equation}
The diagonal elements of the kinetic matrix involve infinite sums but only few contributions to these
sums are relevant since the tunneling of quasiparticles $x$ with high electric charges $q(k-n)$ is suppressed.

The system (\ref{kinetic}) includes an infinite number of equations. It is possible to reduce it to a system of a finite size using a symmetry of the kinetic matrix. Indeed, the matrix elements $M^{a,nq}_{b,mq}$ depend on the final and initial topological charges $b$ and $a$ of edge 2  and on the electric charge of the transmitted quasiparticle $q(n-m)$ but they are independent of the initial electric charge $mq$ of edge 2.
Let us introduce a set of generating functions $p_a(z,t)=\sum_{n}p_{a,nq}z^n$, where the summation extends over all states of edge 2 with the same topological charge $a$.
The kinetic equation (\ref{kinetic}) can be rewritten as

\begin{equation}
\label{kinetic2}
\dot p_a(z)=\sum_b K^a_b(z) p_b(z),
\end{equation}
where $K^a_b(z)=\sum_m M^{a,nq}_{b,mq}z^{n-m}$ and $n$ is arbitrary.
The system (\ref{kinetic2}) contains a finite number of equations.

A general solution of Eq. (\ref{kinetic2}) expresses via the eigenvalues of the matrix $\hat K(z)=K^a_b(z)$.
At $z=1$ the matrix satisfies the conditions of the Rorbach theorem \cite{28}:
All its elements are real, all diagonal elements are negative, all nondiagonal elements are positive or
zero and the sum of the elements in each column is zero. Hence, at $z=1$ its maximal eigenvalue is nondegenerate and equals zero. At $z$ close to one the maximal eigenvalue must be close to zero and still nondegenerate. This strongly simplifies the solution of Eq. (\ref{kinetic2}) at large times $t$. One finds that
$p_{a}(z,t)=p_{a}^{(0)}(z)\exp(\lambda(z)t)$, where $\lambda(z)$ is the maximal eigenvalue of the matrix 
$\hat K(z)$. The conservation of probability means that $\sum_a p_a(z=1,t)=1$. At large $t$ this yields
$\sum_a p_a^{(0)}(z=1)=1$.

We are finally ready to calculate the current (\ref{Q/t},\ref{genQ}). One obtains:

\begin{equation}
\label{QQ}
Q=q\frac{d}{dz}\sum_a p_{a}\big |_{z=1,t}=
q\lambda'(z=1)t\sum_a p_a^{(0)}(z=1)+q\sum_a \frac{d p_a^{(0)}(z=1)}{dz}=qt\lambda'(z=1)
+q\sum_a\frac{d p_a^{(0)}(z=1)}{dz};
\end{equation}

\begin{equation}
\label{II}
I=q\lambda'(z=1).
\end{equation}

Thus, in order to calculate the current, one only needs the eigenvalue $\lambda(z)$ of the matrix $\hat K(z)$. We would like to emphasize again that this matrix is finite. The above equation 
(\ref{II}) can be represented in the computationally more convenient form. Let $\langle \eta|=(1,1,\ldots,1)$,
let 

\begin{equation}
\label{Kv}
\hat K(z) |v(z)\rangle=\lambda(z)|v(z)\rangle
\end{equation} and let $\langle \eta|v\rangle=1$. Note that $\langle\eta|\hat K(1)=0$.
Differentiating Eq. (\ref{Kv}) with respect to $z$ yields: 
$\lambda'|v\rangle+\lambda|v'\rangle=\hat K' |v\rangle +\hat K|v'\rangle$. Multiplying by $\langle\eta|$ and setting $z=1$
produces the final result:

\begin{equation}
\label{AA}
I/q=\lambda'(1)=\langle\eta|\hat A^{(1)}|v\rangle,
\end{equation}
where $\hat A^{(s)}=(zd/dz)^s \hat K(z=1)$.

\section{Finite-temperature noise at low frequency}

The purpose of this section consists in the calculation of the noise at arbitrary temperature for arbitrary statistics.
The results provide an independent check for Section II in the limit of zero temperature. They complement the results for the tunneling probabilities and the current from Sections III and IV. The coefficients $r_{ij}^+$  in the expression (\ref{u-rates}) for the tunneling rates
depend on unknown microscopic parameters and can be extracted by fitting experimental data with the expression for the current, Eq. (\ref{AA}). The same information can be independently obtained from the noise, Eq. (\ref{answer}).

The calculation of the effective charge $e^*$ in the low-frequency limit is based on Eq. (\ref{Fano}).
We already found the average transmitted charge $Q$, Eq. (\ref{QQ}). We need to find 
$\overline{\delta Q^2}=\overline{Q^2}-\bar Q^2$, where $\bar Q=qt\lambda'(1)+q\sum dp^{(0)}_a/dz$, Eq. (\ref{QQ}).
The generating function method yields the following expression for $\overline{Q^2}$:

\begin{equation}
\label{Q2}
\overline{Q^2}=q^2\frac{d}{dz}z\frac{d}{dz}\sum_a p_a\big |_{z=1}=q^2\{[\lambda'(1)]^2t^2+[\lambda''(1)+\lambda'(1)]t
+2\lambda'(1)t\sum_a \frac{d p_a^{(0)}}{dz}+\sum_a \left [\frac{d}{dz}+\frac{d^2}{dz^2}\right ]p_a^{(0)}\}.
\end{equation}

This gives the following answer for the effective charge:

\begin{equation}
\label{e*lambda}
e^*=q[1+\lambda''(1)/\lambda'(1)].
\end{equation}

It is convenient to express the above result via the matrices $A^{(s)}$, Eq. (\ref{AA}).
Differentiating Eq. (\ref{Kv}) twice yields

\begin{equation}
\label{Kv2}
\hat K''|v\rangle+2\hat K'|v'\rangle + \hat K|v''\rangle=\lambda''|v\rangle+2\lambda'|v'\rangle +\lambda|v''\rangle.
\end{equation}
Substituting $z=1$ and multiplying by $\langle \eta|$ gives

\begin{equation}
\label{twoprimes}
\lambda''(1)=\langle\eta|\hat K''|v\rangle+2\langle\eta|\hat K'|v'(1)\rangle=
 \langle\eta|\hat A^{(2)}-\hat A^{(1)} |v\rangle+2\langle\eta|\hat A^{(1)}|v'\rangle,
\end{equation}
where we used the identity $\langle\eta|v'\rangle=[\langle\eta|v\rangle]'=0$.
In order to complete the calculation we need to find $|v'(1)\rangle$. At $z=1$ we can rewrite 
the $z$-derivative of Eq. (\ref{Kv})
in the form $\hat K|v'\rangle=-(K'-\lambda' \hat I)|v\rangle=-(\hat I-|v\rangle\langle\eta| )\hat A^{(1)}|v\rangle$,
where we substituted the expression (\ref{AA}) for $\lambda'$.  Let us now introduce the pseudoinverse matrix
$\tilde K^{-1}$ such that $\hat K\tilde K^{-1}=\tilde K^{-1}\hat K=\hat I-|v\rangle\langle\eta|$, 
$\tilde K^{-1}|v\rangle=0$ and $\langle \eta|\tilde K^{-1}=0$. We find:

\begin{equation}
\label{v'}
|v'(1)\rangle=(\hat I-|v\rangle\langle\eta|)|v'\rangle=-\tilde K^{-1}(\hat I-|v\rangle\langle\eta|)\hat A^{(1)}|v\rangle=-\tilde K^{-1}\hat A^{(1)}|v\rangle.
\end{equation}
Combining equations (\ref{AA},\ref{twoprimes},\ref{v'}) we finally obtain

\begin{equation}
\label{answer}
e^*=q\frac{\langle\eta|\hat A^{(2)}-2\hat A^{(1)}\tilde K^{-1} \hat A^{(1)}|v\rangle}{\langle\eta|\hat A^{(1)}|v\rangle}.
\end{equation}
One can easily check that the above equation reproduces the results of Section II in the Pfaffian and Laughlin cases.
In order to apply the results of Sections IV and V to the Pfaffian state, one has to modify the definition of the topological charge in comparison with Section II. The topological charge can take 6 different values corresponding to 
the 6 vertices of the graph in Fig. 2b. This means that electrons carry the vacuum topological charge. Such a definition is physically sensible since electrons do not acquire statistical phases when they encircle other excitations
\cite{foot-electron}.

\section{Finite-temperature noise at high frequency}

In this section we assume that only one type of quasiparticles can tunnel between the edges. 
We denote the electric charge of the tunneling quasiparticle as $q$ 
and its topological charge as $x$. 
We also assume nondegenerate braiding \cite{kitaev} for the topological charge $x$: This means that for some topological charge $a$, a nontrivial statistical phase $\phi_{ab}\ne 2\pi n$ is accumulated when a particle $x$
makes a circle around $a$. 


If only one type of particles is allowed to tunnel then the expression for the dc-current simplifies:

\begin{equation}
\label{dccurrent}
I=q\sum_{a,b}v_a(w_{a+x\rightarrow b}^+ -w^-_{a\rightarrow b+x}),
\end{equation}
where the tunneling probabilities $w^{\pm}$ can be found in Section III and $v_a$ is the probability to find the topological charge $a$ on edge 2. $v_a$ can be extracted from the stationary solution of the kinetic equation (\ref{kinetic})
and are the components of the vector $|v\rangle$, Eq. (\ref{Kv}).

The expression for the noise (\ref{noise})
can be rewritten as

\begin{equation}
\label{hfn}
S=\frac{1}{4\Delta t}\int_{-\Delta t}^{\Delta t}dt_1\int_{-\infty}^{\infty}dt_2
\langle I(t_1)I(t_2)+I(t_2)I(t_1)\rangle\exp(i\omega [t_2-t_1]),
\end{equation}
where the time $\Delta t$ is much greater than the duration of one tunneling event and much shorter than the inverse frequency $1/\omega$. Contributions from large times $|t_2|>1/\omega$ are suppressed by the rapidly oscillating exponential factor. Contributions from the region $\Delta t<|t_2|<1/\omega$ correspond to rare situations when one tunneling event occurs at the moment of time $t_1$, $|t_1|<\Delta t$, and another one at the moment of time $t_2$,
$|t_2|<1/\omega$. Neglecting them, we can reduce Eq. (\ref{hfn}) to the following form:

\begin{eqnarray}
\label{hfnQ}
S=\frac{1}{4\Delta t}\int_{-\Delta t}^{\Delta t}dt_1\int_{-\Delta t}^{\Delta t}dt_2\langle I(t_1)I(t_2)+I(t_2)I(t_1)\rangle= & & \nonumber \\
\frac{1}{4\Delta t}\int_{-\Delta t}^{\Delta t}dt_1\int_{-\Delta t}^{\Delta t}dt_2\left\langle \frac{dQ(t_1)}{dt}
\frac{dQ(t_2)}{dt}+\frac{dQ(t_2)}{dt}\frac{dQ(t_1)}{dt}\right\rangle=\frac{1}{2\Delta t}
\langle[Q(\Delta t)-Q(-\Delta t)]^2\rangle=
\frac{1}{2\Delta t}\langle\Delta Q^2\rangle, & &
\end{eqnarray}
where $\Delta Q$ denotes the change of the electric charge of edge 2 after the time interval $2\Delta t$ and we used the relations $\exp(i\omega\Delta t)\approx 1$ and $I=dQ/dt$. Possible values of $\Delta Q=nq$ correspond to an integer number
of transmitted quasiparticles. Neglecting the contributions from large $n$, $|n|>1$, one finds

\begin{equation}
\label{noise-inter}
S=q^2\sum_{a,b}v_a(w_{a+x\rightarrow b}^+ +w^-_{a\rightarrow b+x}).
\end{equation}
This expression can be evaluated on the basis of the kinetic equations discussed in section IV.

The situation considerably simplifies in the thermal equilibrium case, $V=0$, since the distribution function $v_a$
can be found from the detailed balance principle. At zero voltage the tunneling rates $u^{\pm}$, 
Eq. (\ref{dbc}), are equal and hence $d_a^2w_{a+x\rightarrow b}^+ =d_b^2w^-_{b\rightarrow a+x}$. From the detailed
balance condition $v_a w_{a+x\rightarrow b}^+ =v_b w^-_{b\rightarrow a+x}$, we get

\begin{equation}
\label{dbv}
v_a=\frac{d_a^2}{D^2},
\end{equation}
where the normalization constant is the total quantum dimension $D=\sqrt{\sum_a d_a^2}$.
The substitution of (\ref{dbv},\ref{w+},\ref{w-}) into Eq. (\ref{noise-inter}) yields

\begin{equation}
\label{S}
S=S_1+S_2+S_2^*,
\end{equation}
where $S_2$ depends on the magnetic flux and $S_1$ exhibits no flux dependence:

\begin{equation}
\label{S1}
S_1=\frac{2q^2}{d_xD^2}r_{11}^+\sum_{a,b}N^b_{ax}d_a d_b[|\Gamma_1|^2+|\Gamma_2|^2];
\end{equation}

\begin{equation}
\label{S2}
S_2=\frac{2q^2}{d_xD^2}\sum_{ab}N^b_{ax}d_a d_b \frac{\theta_b}{\theta_a\theta_x}r_{12}^+\Gamma_1^*\Gamma_2\exp(i\phi_{\rm AB}).
\end{equation}
The expression for $S_1$ can be simplified with the identity $\sum_b N^b_{ax}d_b=d_ad_x$, Ref. \onlinecite{kitaev}.
This yields

\begin{equation}
\label{S1-final}
S_1=2q^2r_{11}^+[|\Gamma_1|^2+|\Gamma_2|^2].
\end{equation}
It is not surprising that the flux-independent noise expresses via the fractional charge and contains no information about fractional statistics. Interestingly, the same is true for the total noise $S$. In fact, $S=S_1$ and
 $S_2=0$ as can be seen from the following argument. 
The combination $\frac{1}{D}\sum_b N^b_{ax}d_b\frac{\theta_b}{\theta_a\theta_x}=s_{a\bar x}=s^*_{xa}$ is an element of the
topological $S$-matrix \cite{kitaev}. Hence, $S_2=2q^2 r_{12}^+\Gamma_1^*\Gamma_2\exp(i\phi_{\rm AB})\sum_a  s^*_{xa}(d_a/D)=
{\rm const}\sum_a  s^*_{xa}s_{1a}=0$. In the last equality we used the orthogonality of the nontrivial fusion character
$s_{xa}/s_{x1}$ and the trivial character $s_{1a}/s_{11}$. Thus, the equilibrium high-frequency noise is independent of fractional statistics and the magnetic flux through the interferometer and does not exhibit quantum interference effects
in the lowest order in $\Gamma$'s. We would like to emphasize our assumption that the tunneling quasiparticle exhibits nontrivial braiding with other
topological charges. Otherwise, $S_2\ne 0$. In particular, this means that
quantum interference can be observed, if electron tunneling is allowed. Note that the equilibrium
high-frequency noise (\ref{S},\ref{S1-final}) is related to the linear conductance by the fluctuation-dissipation theorem.

\section{Conclusions}

In conclusion, we have calculated shot noise in the Mach-Zehnder
interferometer in fractional QHE states. At zero temperature, the Fano factor is always
smaller than one in Abelian states and can exceed one in the Pfaffian state. 
Thus, noise in the Mach-Zehnder interferometer can be used for an experimental probe
of non-Abelian statistics. An advantage of such a setup consists in the robustness with respect to the fluctuations of the number of the quasiparticles trapped inside the system.
Our predictions are independent of the model of the edges. They apply even in the regime
of ``short edges", where the physics of the edges cannot be separated from the physics of the leads in the relevant energy window. The only assumption about the edges consists in their chirality.

We also obtained general expressions for the current and noise at arbitrary temperatures and voltages for any anyonic statistics. The expressions drastically simplify at zero voltage in the high-frequency regime. In that
regime no quantum interference effects can be observed for any fractional statistics in sharp contrast with the case of fermions. 

Our analysis assumes weak quasiparticles tunneling. For Fermi and Bose systems this would mean a single-particle problem. For anyons, many-particle effects persist even for weak tunneling due to the ``black hole'' in the center of the interferometer (the hole encircled by edge 2 in Fig. 1). The ``black hole'' is entangled with the rest of the system and interacts with tunneling topological charges. This results in rather unusual behavior including non-analytic dependence of the current and noise on the small tunneling amplitudes. It should be possible to extend our approach
to more complicated geometries with several ``black holes''.

\begin{acknowledgments}

We acknowledge the support by ARO under Grants No.\
W911NF-04-1-0236 (A.\,K.) and W911NF-05-1-0294 (A.\,K.), by NSF
under Grants  No.\ PHY99-07949 (D.\,E.\,F.), PHY-0456720 (A.\,K.)
and DMR-0544116 (D.\,E.\,F. and K.\,T.\,L.), by the U.S.-Israel
BSF (Y.\,G. and A.\,S.), the Minerva Foundation (A.\,S.) and the
ISF of the Israel Academy of Sciences (Y.\,G.). D.\,E.\,F.
acknowledges the hospitality of Microsoft Station Q and KITP in
Santa Barbara.

\end{acknowledgments}

\begin{figure}
\epsfig{file=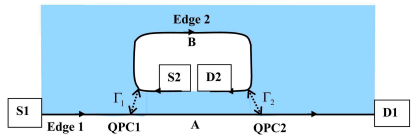, scale=0.4} \caption{Schematic picture of the
anyonic Mach-Zehnder interferometer. Arrows indicate the edge mode propagation
directions from source S1 to drain D1 and from source S2 to drain D2.} \label{fig1}
\end{figure}

\begin{figure}
\epsfig{file=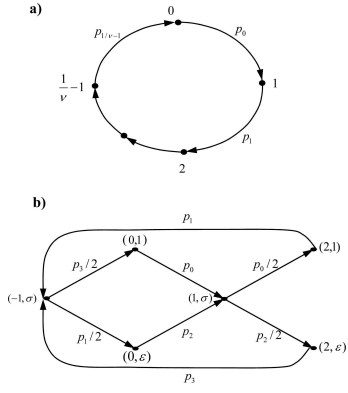, scale=0.4} \caption{a) States and transition
probabilities in a Laughlin liquid with the filling factor $\nu$.
b) In the $\nu=5/2$ QHE liquid, each state of the interferometer
is labeled by  the topological charge of edge 2 and the number of
the previous tunneling events ${\rm mod}~4$.} \label{fig2}
\end{figure}


\begin{figure}
\epsfig{file=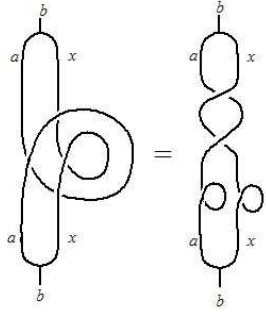, scale=0.4}\caption{The diagram for the calculation of the statistical phase $\phi_{ab}$.
Note that the left diagram depends only on the combined topological charge $b$ of the excitations $a$ and $x$ which should be viewed as one composite object at the calculation of that diagram.}  
\label{fig3}
\end{figure}


\begin{figure}
\epsfig{file=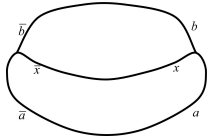, scale=0.4} \caption{The diagram for the calculation of the fusion amplitudes.} \label{fig4}
\end{figure}


\begin{thebibliography} {150}

\bibitem{MooreRead} G.\,Moore and N.\,Read,
Nucl. Phys. B \textbf{360}, 362 (1991).

\bibitem{nw}
C. Nayak and F. Wilczek, Nucl. Phys. B {\bf 479}, 529 (1996).

\bibitem{Kitaev97} A.\,Yu.\,Kitaev,
Annals of Physics \textbf{303}, 2 (2003).

\bibitem{kitaev}
A. Kitaev, Annals of Physics {\bf 321}, 2 (2006).


\bibitem{weizmann-G} R. de Picciotto, M. Reznikov, M. Heiblum, V. Umansky, G. Bunin, and D. Mahalu, 
Nature {\bf 389}, 162
(1997); L. Saminadayar, D. C. Glattli, Y. Jin, and B. Etienne, Phys. Rev. Lett.  {\bf 79},
2526 (1997).

\bibitem{cfksw} C. de C. Chamon, D. E. Freed, S. A. Kivelson, S. L. Sondhi, and X.  G. Wen, 
Phys. Rev. B {\bf 55}, 2331
(1997).

\bibitem{HBT}
S. B. Isakov, T. Martin, and S. Ouvry, Phys. Rev. Lett. {\bf 83},
580 (1999); I. Safi, P. Devillard, and T. Martin, {\it ibid.} {\bf
86}, 4628 (2001); S. Vishveshwara, {\it ibid.} {\bf 91}, 196803
(2003); E.-A. Kim, M. Lawler, S. Vishveshwara, and E. Fradkin, {\it ibid.} {\bf 95}, 176402
(2005).

\bibitem{kane}
C. L. Kane, Phys. Rev. Lett. {\bf 90}, 226802 (2003).

\bibitem{lfg}
K. T. Law, D. E. Feldman, and Y. Gefen, Phys. Rev. B {\bf 74},
045319 (2006).

\bibitem{gss}
E. Grosfeld, S. H. Simon, and A. Stern, Phys. Rev. Lett. {\bf 96},
226803 (2006).

\bibitem{FNTW}
E. Fradkin, C. Nayak, A. Tsvelik, and F. Wilczek, Nucl. Phys. {\bf B516}, 704 (1998).

\bibitem{DFN}
S. Das Sarma, M. Freedman, and C. Nayak, Phys. Rev. Lett. {\bf 94}, 166802 (2005).

\bibitem{BKS}
A. Stern and B. I. Halperin, Phys. Rev. Lett. {\bf 96}, 016802
(2006).

\bibitem{SH}
P. Bonderson, A. Kitaev, and K. Shtengel, Phys. Rev. Lett. {\bf
96}, 016803 (2006).

\bibitem{chamon}
C.-Y. Hou and C. Chamon, Phys. Rev. Lett. {\bf 97}, 146802 (2006).

\bibitem{fk}
D. E. Feldman and A. Kitaev, Phys. Rev. Lett. {\bf 97}, 186803 (2006).

\bibitem{Goldman} F. E. Camino, W. Zhou, and V. J. Goldman
Phys. Rev. B {\bf 74}, 115301 (2006) and references therein;
cond-mat/0611443.

\bibitem{Rosenow} B. Rosenow and B. I. Halperin, Phys. Rev. Lett. {\bf 98}, 106801 (2007).

\bibitem{MZ}
Y. Ji,  Y. C. Chung, D. Sprinzak, M. Heiblum, D. Mahalu, and H. Shtrikman, Nature {\bf 422}, 415 (2003).

\bibitem{newreview}
S. Das Sarma, M. Freedman, C. Nayak, S. H. Simon, and A. Stern, arXiv:0707.1889.

\bibitem{footnote}
Note the coefficient $1/2$ in our definition.

\bibitem{mznoise}
F. Marquardt and C. Bruder, Phys. Rev. Lett. {\bf 92}, 056805
(2004); F. Marquardt, Europhys. Lett. {\bf 72}, 788 (2005); V. S.
W. Chung, P. Samuelsson, and M. B\"{u}ttiker, Phys. Rev. B {\bf
72}, 125320 (2005); H. Forster, S. Pilgram, and M. B\"{u}ttiker,
{\it ibid.}, 075301 (2005). M. Veillette, J. Chalker and Y. Gefen,
cond-mat/0703162.

\bibitem{kf94}
C. L. Kane and M. P. A. Fisher, Phys. Rev. Lett. {\bf 72}, 724
(1994).

\bibitem{foot1}
At $\nu=1/3$ the quasi-hole transfer operator is the only relevant operator \cite{rev}.
At $\nu=5/2$ the operator which transfers electric charge $e/2$ and topological charge 1
is relevant; the operator which transfers zero electric charge and topological charge $\epsilon$
is marginal\cite{ffn}.

\bibitem{rev}
C. L. Kane and M. P. A. Fisher, in {\it Perspectives in Quantum Hall Effect}, edited by S. Das Sarma
and A. Pinczuk (John Wiley \& Sons, New York, 1997).

\bibitem{ffn}
P. Fendley, M. P. A. Fisher, and C. Nayak,
Phys. Rev. B {\bf 75}, 045317 (2007).


\bibitem{28}
E. B. Fel'dman, Theoretical and Experimental Chemistry {\bf 10}, 645 (1976).

\bibitem{foot-electron}
Indeed, any anyonic system is in fact made of electrons. When one electron makes a full circle around any number of other electrons, only the phase of the form $2\pi n$ can be accumulated. This is equivalent to no phase at all. 



\end{thebibliography}
\end{document}